\documentstyle[12pt]{article}
\newcommand{\beq}{\begin{equation}}
\newcommand{\eeq}{\end{equation}}
\newcommand{\bea}{\begin{eqnarray}}
\newcommand{\eea}{\end{eqnarray}}

\font\rmtit=cmr12
\font\myit=cmti10 scaled\magstep1

\begin{document}
\title{
\hfill {\rmtit DFTT 26/1999}\\
\vspace{-0.3cm}\hfill {\rmtit UWThPh-1999-31}\\
%\vspace*{1.cm}
{\bf New developments in neutrino physics}}
\author{
W.M. Alberico$^{\mathrm{a}}$,
S.M. Bilenky$^{\mathrm{b}}$,
\vspace*{0.3cm}\\
%\begin{tabular}{c}
{\myit $^{\mathrm{a}}$INFN, Sezione di Torino}\\
{\myit and Dipartimento di Fisica Teorica, Universit\`a di Torino,}
\\
{\myit Via P. Giuria 1, 10125 Torino, Italy}
\\
{\myit $^{\mathrm{b}}$Joint Institute for Nuclear Research, Dubna, Russia}
\\
{\myit and Institute for Theoretical Physics, University of Vienna}\\
{\myit Boltzmanngasse 5, A--1090 Vienna, Austria}
}
%\end{tabular}

\date{\today}
\maketitle
\begin{abstract}
A  review of the problem of neutrino mass, mixing and oscillations
is given. Possible phenomenological schemes of neutrino mixing are
discussed. The most important consequences of neutrino mixing--neutrino
oscillations are considered in some details. The data of
atmospheric, solar and LSND experiments are discussed. The results of
phenomenological analyses of the data under the assumption of the mixing
of three and four massive neutrinos are shortly presented.
\end{abstract}

\section{Introduction}

The strong evidence in favour of the oscillations of atmospheric
neutrinos, which was recently obtained by the Super--Ka\-mio\-kan\-de
collaboration\cite{SKamio98}, has attracted the attention of many
physicists on neutrinos. The problem of masses and mixing of
neutrinos is at present the central problem of elementary particle
physics. The investigation of this problem is one of the major tools
of searching for new physics beyond the Standard Model.

Indications in favour of neutrino oscillations were obtained in all
modern atmospheric neutrino experiments (IMB\cite{IMB},
Soudan-2\cite{Soudan}, Ka\-mio\-kan\-de\cite{Kamioka},
Super-Ka\-mio\-kan\-de\cite{SKamio98}), in all solar neutrino experiments
(Homestake\cite{Homesta}, GALLEX\cite{Gallex}, SAGE\cite{Sage},
Ka\-mio\-kan\-de\cite{Kamioka2}, Super-Ka\-mio\-kan\-de\cite{Skamioka},
MACRO\cite{MACRO}) and in the accelerator LSND experiment\cite{LSND}.

From all existing data it follows that there are three different scales
of neutrino mass squared differences, $\Delta m^2$:
$\Delta m^2_{\mathrm solar}\simeq 10^{-5}$~eV$^2$ (or $10^{-10}$~eV$^2$),
$\Delta m^2_{\mathrm atm}\simeq 10^{-3}$~eV$^2$ and
$\Delta m^2_{\mathrm LSND}\simeq 1$~eV$^2$. This implies that at least
four massive neutrinos exist in nature, i.e. the number of massive
neutrinos is larger than the number of flavour neutrinos ($\nu_e,\,
\nu_{\mu},\, \nu_{\tau}$). Thus, if future neutrino oscillation experiments
will confirm the existing results, it will imply that neutrino mixing
and quark mixing are of a different origin.

Many new neutrino oscillation experiments are in preparation. The region
of $\Delta m^2$ of atmospheric neutrinos ($\Delta m^2_{\mathrm atm}$ )
will be investigated in the long--baseline (LBL)
experiments MINOS\cite{Minos}, ICARUS\cite{Icarus}, OPERA\cite{Opera} et
al., the solar neutrino region ($\Delta m^2_{\mathrm solar}$) in SNO\cite{SNO},
BOREXINO\cite{Borexino} and in the reactor LBL experiment
KAMLAND\cite{Kamland}.
Finally, the LSND region of $\Delta m^2$ ($\Delta m^2_{\mathrm LSND}$)
 will be investigated in the
future short--baseline (SBL) experiment BOONE\cite{Boone}.

Neutrino masses, mixing and nature (Dirac or Majorana) are of fundamental
importance for the theory. It is common belief (see ref.\cite{sam1}) that
neutrino masses and mixing are generated by a mechanism beyond the Standard
Model. The main reason for that is the experimental fact that neutrino masses
are much smaller than the masses of all the other fundamental fermions
(leptons and quarks).
The full understanding of the origin of neutrino masses and mixing
will require, however, many new experiments.

Neutrinos are very important in astrophysics: massive neutrinos are
plausible candidates for hot dark matter particles, the number of
neutrino species plays a crucial role in the Big Bang Nucleosynthesis,
and so on. We will not discuss here these relevant issues.

In section 2 we will consider the general phenomenological framework for
neutrino masses and mixing. In section 3 we will discuss  neutrino
oscillations. In section 4 the latest experimental data will be examined.
Finally in section 5 the analysis of the data will be presented.

\section{Neutrino Mixing}

The neutrinos $\nu_e, \nu_{\mu}, \nu_{\tau}$, 
which are produced in weak processes like pion and muon decays,
nuclear beta decays etc. are called {\it flavour neutrinos}.
In the interaction with nucleons a flavour neutrino $\nu_{\ell}$
($\ell=e,\mu,\tau$) produces the lepton $\ell^-$ and hadrons (CC) or the
same neutrino $\nu_{\ell}$ and hadrons (NC).
From all the available data it follows that the interaction of flavour
neutrinos is perfectly described by the Lagrangian of the Standard
Model:
\beq
{\cal L}_I = \left( -\frac{g}{2\sqrt{2}}j_{\alpha}^{CC} W^{\alpha}
+ h.c.\right) -\frac{g}{2\cos\theta_W} j_{\alpha}^{NC} Z^{\alpha}.
\label{Lagr}
\eeq
Here the charged ($j_{\alpha}^{CC}$) and neutral ($j_{\alpha}^{NC}$)
currents are given by:
\bea
j_{\alpha}^{CC}&=& 2\sum_{\ell=e,\mu,\tau}
{\bar\nu}_{\ell L}\gamma_{\alpha} \ell_L +\dots\, ,
\label{Ccurr}\\
j_{\alpha}^{NC}&=& \sum_{\ell=e,\mu,\tau}
{\bar\nu}_{\ell L}\gamma_{\alpha} \nu_{\ell L} +\dots\, ;
\label{Ncurr}
\eea
$W^{\alpha}$ and $Z^{\alpha}$ are the fields of the vector bosons
$W^{\pm}$ and $Z^0$, $\theta_W$ is the weak mixing angle and $g$ is the
coupling constant. From LEP data it follows that the number of light
flavour neutrinos, $n_{\nu_f}$, is equal to three\cite{LEP}:
\beq
 n_{\nu_f} = 2.994 \pm 0.012
 \label{nnu}
\eeq

The Lagrangian (\ref{Lagr}) conserves the additive electron $L_e$,
muon $L_{\mu}$ and tauon $L_{\tau}$ lepton numbers:
\beq
\sum L_e={\mathrm const},\qquad
\sum L_{\mu}={\mathrm const},\qquad
\sum L_{\tau}={\mathrm const}\, .
\eeq
According to the {\it neutrino mixing hypothesis} this law is an
approximate one: it is violated by the {\it neutrino mass
term}.

The neutrino mass term (see the review article\cite{CSW} and references 
therein) can be completely different from the corresponding
lepton and quark mass terms. This is connected with the fact that
neutrinos with definite masses can be {\it Dirac or Majorana
particles}; charged leptons and quarks are {\it Dirac particles}.
The neutrino mass term can be written in the following,
general form:
\beq
{\cal L} = - {\bar n}_R M n_L + h.c.
\label{numass}
\eeq
where $n_{L,R}$ are columns of the neutrino fields and $M$ is a matrix.
There are two general possibilities for $n_L$.

{\bf I.} The column $n_L$ contains only flavour neutrino fields:
\beq
 n_L=\left(\begin{array}{c}
\nu_{e L}\\ \nu_{\mu L}\\ \nu_{\tau L}\end{array}\right)
\label{vector1}
\eeq
In this case $M$ is a $3\times 3$ matrix and for the mixing we have
\beq
\nu_{\ell L}=\sum_{i=1}^3 U_{\ell i}\nu_{iL}\qquad(\ell=e,\mu,\tau)
\label{mixing1}
\eeq
where $U^{\dagger}U=1$ and $\nu_i$ is the field of neutrinos with mass
$m_i$. Only transitions between flavour neutrinos 
$\nu_{\ell}\rightleftharpoons\nu_{\ell'}$ are possible in this case.

The nature of $\nu_i$ depends on $n_R$: if
\beq
 n_R=\left(\begin{array}{c}
\nu_{e R}\\ \nu_{\mu R}\\ \nu_{\tau R}\end{array}\right)\, ,
\label{vector1Ra}
\eeq
where $\nu_{\ell R}$ are right--handed neutrino fields,
global gauge invariance
\beq
\nu_{\ell L}\to e^{i\alpha}\nu_{\ell L}, 
\nu_{\ell R}\to e^{i\alpha}\nu_{\ell R}, 
\ell \to e^{i\alpha} \ell 
\label{gauge}
\eeq
($\alpha$ being a real constant, the same for all fields) takes place. 
Hence, in this case the total lepton charge
\beq
L= L_e + L_{\mu} + L_{\tau}
\label{leptch}
\eeq
is conserved and  fields of neutrinos  with definite masses, $\nu_i$,
are {\it Dirac fields} ($\nu_i$ and the charge conjugated field, 
$\nu_i^C=C{\bar\nu}_i^T$ are independent). The corresponding mass term 
is called Dirac mass term.
Let us notice that Dirac mass term can be generated in the framework of 
the standard Higgs mechanism which is responsible for the generation of 
the masses of charged leptons and quarks.

If, instead, 
\beq
 n_R=\left(\begin{array}{c}
(\nu_{e L})^C \\ (\nu_{\mu L})^C \\ (\nu_{\tau L})^C 
\end{array}\right)\, ,
\label{vector1Rb}
\eeq
where $(\nu_{\ell L})^C=C{\bar\nu}_{\ell L}^T$, is the right--handed 
component, then there are no conserved lepton numbers and the fields 
of neutrinos with definite masses are {\it Majorana fields}
($\nu_i^C=\nu_i$).\footnote{For Majorana neutrino not only the 
electric charge, but also all lepton charges are equal to zero}  
The corresponding mass term is called Majorana mass term.

Let us notice that if massive neutrinos are Majorana particles, a 
process like neutrinoless double beta decay
\[ (A,Z) \to (A,Z+2) + e^- + e^-, \]
in which the total lepton number is not conserved, becomes possible,
There is no difference in neutrino oscillations for the case of 
Dirac or Majorana masses.

{\bf II.} In the most general case not only the three flavour neutrino 
fields enter into $n_L$, but also other fields $\nu_{s L}$ ($s=s_1,\dots$), 
which are not contained in the Standard Lagrangian of weak interactions, 
eq.(\ref{Lagr}), and hence are called {\it sterile fields}:
\beq
 n_L=\left(\begin{array}{c}
\nu_{e L}\\ \nu_{\mu L}\\ \nu_{\tau L}\\ \nu_{s_1 L}\\ \vdots
\end{array}\right)
\label{vector2}
\eeq
Sterile fields can be right--handed neutrino fields 
($\nu_{sL}=(\nu_{sR})^C$ ) and/or fields of SUSY particles\footnote{
If the $\nu_{sR}$ fields enter only into the neutrino mass term, then 
the corresponding particles are really sterile. Should there be a 
right--handed interaction, then ``sterile'' particles could experience a
much weaker interaction than the standard electroweak interaction.}

For the mixing we have, in this case
\beq
\nu_{\alpha L}=\sum_{i=1}^{3+n_s} U_{\alpha i}\nu_{iL}
\qquad (\alpha=e,\mu,\tau,s_1,\dots)
\label{mixing2}
\eeq
where $\nu_i$ is the neutrino field with mass $m_i$ and $U$ is
a $(3+n_s)\times(3+n_s)$ unitary mixing matrix. The number of sterile 
fields, $n_s$, can only be fixed by a model. For $\nu_{sL}=(\nu_{sR})^C$
it is natural to assume that $n_s=3$. 

If the neutrino masses $m_i$ are small ($i=1,\dots, 3+n_s$), then not
only oscillations between flavour neutrinos 
$\nu_{\ell}\rightleftharpoons\nu_{\ell'}$, but also oscillations between 
flavour and sterile neutrinos $\nu_{\ell}\rightleftharpoons\nu_s$ will
take place.

The nature of massive neutrinos $\nu_i$ depends on $n_R$. If 
$n_R=(n_L)^C$ neutrinos $\nu_i$ are Majorana particles and 
neutrinoless double $\beta$--decay is possible. The corresponding
mass term is called Dirac--Majorana mass term.

Majorana neutrino masses 
can be generated only in the framework of models beyond the Standard
Model. In the case of the Dirac--Majorana mass term there exists  
a plausible (the so called see--saw) mechanism for neutrino 
mass generation\cite{sam3}.
It is based on the assumption that lepton numbers are violated by 
the right--handed Majorana mass term at a scale $M$ much larger than 
the electroweak scale. 

The spectrum of masses of Majorana particles in the see--saw case 
contains three light neutrinos $\nu_i$ with masses $m_i$ and three 
very heavy Majorana particles with masses $M_i\simeq M$. The two set of 
masses are connected by the see--saw relation
\beq
m_i\simeq \frac{(m_f^i)^2}{M_i} \ll m_f^i\, ,
\qquad\quad (i=1,2,3)
\label{seesaw}
\eeq
where $m_f^i$ is the mass of a quark or a lepton in the ``i-th'' family.
The see--saw mechanism connects the smallness of Majorana neutrino masses
with the violation of lepton numbers at very large mass scales. Notice 
that in the see--saw case the neutrino masses satisfy a hierarchy relation:
\beq
m_1 \ll m_2 \ll m_3\, .
\label{hierar}
\eeq

\section{Neutrino oscillations}

In this section we will discuss the phenomenon of oscillations in
neutrino beams\cite{sam4} which can occur if neutrino masses are
different from zero and flavour neutrino fields are mixtures of
massive fields [see (\ref{mixing1}) and (\ref{mixing2})]. In this
case for a state with momentum ${\vec p}$ we have
\beq
|\nu_{\alpha}>=\sum_i U_{\alpha i}^*|i>
\qquad(\alpha=e,\mu,\tau,s_1,\dots),
\label{nualfa}
\eeq
where $|i>$ is the state of a neutrino with mass $m_i$, momentum
${\vec p}$ and energy
\beq
E_i=\sqrt{p^2+m_i^2} \simeq p+\frac{m_i^2}{2p}
\qquad (p \gg m_i).
\label{energy}
\eeq
The state $|i>$ is eigenstate of the free Hamiltonian $H_0$:
\beq
H_0 |i> = E_i|i>\, .
\label{schrod}
\eeq
Relation (\ref{nualfa}) implies that the states of flavour
neutrinos (and eventually the sterile ones) are {\it coherent
superpositions} of neutrino states with different masses.
This is valid only when the neutrino mass differences are
small and, due to the uncertainty principle, different mass
components cannot be distinguished in production and detection
processes (for a recent discussion of this problem see
ref.\cite{sam5}).

If at $t=0$ the state of neutrinos is $|\nu_{\alpha}>$, the
probability of the transition into the state $|\nu_{\beta}>$
after a time $t$ is given by:
\bea
<\nu_{\beta}|e^{-iH_0 t}|\nu_{\alpha}> &=&
\sum_i<\nu_{\beta}|i> e^{-iE_i t} <i|\nu_{\alpha}>
\nonumber \\
&=& \sum_i U_{\beta i} e^{-iE_i t} U_{\alpha i}^*
\label{prob}
\eea
From (\ref{prob}), using the unitarity of the mixing matrix $U$, the 
following general expression for the probability of the transition
$\nu_{\alpha}\to\nu_{\beta}$ can be obtained:
\beq
P(\nu_{\alpha}\to\nu_{\beta}) =
\left| \delta_{\beta\alpha} +\sum_i U_{\beta i}
\left(e^{-i\Delta m_{i1}^2\frac{L}{2p}} -1\right)
U_{\alpha i}^*\right|^2\, .
\label{prob1}
\eeq
Here $L\simeq t$ is the distance between the neutrino source
and the neutrino detector, $\Delta m_{i1}^2=m_i^2-m_1^2$ (we
have assumed that $m_1<m_2<\dots$). The  probability
for an antineutrino transition
${\bar\nu}_{\alpha}\to {\bar\nu}_{\beta}$ is given by
\beq
P({\bar\nu}_{\alpha}\to {\bar\nu}_{\beta}) =
\left| \delta_{\beta\alpha} +\sum_i U_{\beta i}^*
\left(e^{-i\Delta m_{i1}^2\frac{L}{2p}} -1\right)
U_{\alpha i}\right|^2\, .
\label{prob2}
\eeq
Obviously, as a consequence of CPT invariance,
\beq
P(\nu_{\alpha}\to\nu_{\beta}) =
P({\bar\nu}_{\beta}\to {\bar\nu}_{\alpha}) .
\eeq
Let us notice that from CP invariance  it follows:
\beq
P(\nu_{\alpha}\to\nu_{\beta}) =
P({\bar\nu}_{\alpha}\to {\bar\nu}_{\beta}) .
\eeq

Thus the probability of the transition
$\nu_{\alpha}\to\nu_{\beta}$
(${\bar\nu}_{\alpha}\to{\bar\nu}_{\beta}$) depends, in the
general case, on $n-1$ neutrino mass squared differences
($n=3+n_s$) and on the parameter $L/p$. When, for all values
of $i$, $\Delta m_{i1}^2\ll p/L$ then neutrino oscillations
cannot be observed [$P(\nu_{\alpha}\to\nu_{\beta})=
\delta_{\alpha\beta}$]. In order to observe neutrino
oscillations is it necessary that for some value of $\kappa\ge 2$, 
$\Delta m_{\kappa i}^2\ge p/L$.

Let us consider the simplest case of the mixing of two
neutrino species. Then
\beq
U=\left(\begin{array}{cc} \cos\theta& \sin\theta\\
-\sin\theta &\cos \theta\end{array}\right)
\label{U2dim}
\eeq
where $\theta$ is the mixing angle. From (\ref{prob1}) and
(\ref{prob2}) one gets
\bea
P(\nu_{\alpha}\to\nu_{\beta}) &&=
P({\bar\nu}_{\alpha}\to {\bar\nu}_{\beta})
\nonumber \\
&&\qquad = \frac{1}{2}\sin^2{2\theta}
\left(1-\cos\frac{\Delta m^2 L}{2p}\right)
\quad (\beta\ne\alpha)
\label{trans1}\\
P(\nu_{\alpha}\to\nu_{\alpha})&&=
P(\nu_{\beta}\to\nu_{\beta}) =
1- P(\nu_{\alpha}\to\nu_{\beta})                    
\nonumber \\
&&\qquad = 1- 
\frac{1}{2}\sin^2{2\theta}
\left(1-\cos\frac{\Delta m^2 L}{2p}\right)\, .
\label{trans2}
\eea
In the above $\Delta m^2=m_2^2 - m_1^2$ and 
$\alpha,\beta$ can assume the values $e,\mu$ or $\mu,\tau$ and so on.
The expressions (\ref{trans1}), (\ref{trans2}) are written
in units $\hbar=c=1$. 
The transition probability can also be written in the form:
\beq
P(\nu_{\alpha}\to\nu_{\beta}) =\frac{1}{2}\sin^2{2\theta}
\left(1-\cos{2.54}\frac{\Delta m^2 L}{E}\right)
\label{trans1bis}
\eeq
where $L$ is the distance in m, $E$ is the neutrino energy in
MeV, $\Delta m^2$ is the neutrino mass--squared difference in eV$^2$. 
For the oscillation length we have, from (\ref{trans1}) and 
(\ref{trans1bis}),
\beq
L_0=4\pi\frac{E}{\Delta m^2}= 2.47\frac{E({\mathrm MeV})}
{\Delta m^2({\mathrm eV}^2)}\,{\mathrm m}\, .
\label{length}
\eeq

In the simplest case of two neutrinos mixing, the probability
of neutrino transitions depends on the parameters $\sin^2{2\theta}$
(the amplitude of the oscillations) and $\Delta m^2$, which
characterizes the oscillation length.
The necessary condition for the oscillations to be observable is
\beq
\Delta m^2\frac{L}{E}\ge 1\, .
\label{observability}
\eeq
Thus, the larger the value of the parameter $L/E$, the more sensitive
will be an experiment to the value of $\Delta m^2$.
Typical values of the parameter $L$(m)/$E$(MeV) for SBL and LBL
accelerator experiments, for SBL and LBL reactor experiments, for
atmospheric neutrino experiments and for solar neutrino experiments
are
\[ 1; 10^2\div 10^3; 10^2; 10^3; 10^2\div 10^3; 10^{11},\]
respectively.

\section{The status of neutrino oscillations}

Evidences and indications in favour of neutrino oscillations were
found in many neutrino oscillation experiments. We will discuss
here shortly the results that have been obtained.

\subsection{Atmospheric neutrinos}
Atmospheric neutrinos are mainly produced in the decay processes
\beq
\pi^{\mp}\to\mu^{\mp} \nu_{\mu}({\bar\nu}_\mu),
\qquad \mu^{\mp}\to e^{\mp}{\bar\nu_e}\nu_{\mu}(\nu_e{\bar\nu}_\mu),
\label{pidecay}
\eeq
pions being produced in the interaction of cosmic rays with nuclei in
the Earth atmosphere. At energies $\le 3$~GeV the ratio of fluxes of
$\nu_{\mu}, {\bar\nu}_{\mu}$ and $\nu_e, {\bar\nu}_e$ is equal to
two, while at higher energies it is larger than two (since
not all muons have time to decay in the atmosphere). The ratio can be
predicted, however, with accuracy better than $ 5\%$ (the absolute fluxes
of muon and electron neutrinos cannot be calculated with accuracy
better than $20\%$).

The results of atmospheric neutrinos experiments are usually presented in the
form of a double ratio:
\beq
R=\left(\frac{N_\mu}{N_e}\right)_{exp}/
\left(\frac{N_\mu}{N_e}\right)_{MC}
\label{ratio}
\eeq
where $N_\mu(N_e)$ is the total number of muon (electron) events 
(in modern detectors neutrino and antineutrino events cannot be 
distinguished) and $\left(N_\mu/N_e\right)_{MC}$ is the ratio predicted 
by Monte Carlo simulations.

We shall  discuss mainly the results of the Super--Kamio\-kan\-de
experiment\cite{SKamio98}. In this experiment a large 50~Ktons
water Cherenkov detector is used. The detector consists of two parts:
the inner one (22.5~Ktons fiducial volume) is covered with 11146
photomultipliers; the outer part, 2.75~m thick, is covered with
1885 photomultipliers. The electrons and muons are detected by
observing their Cherenkov radiation. The effectiveness of particle
identification is larger than 98~\%.

The observed events are divided in fully contained events (FC),
for which all Cherenkov light is deposited in the inner detector,
and partially contained  events (PC), in which the muon track
deposits part of its Cherenkov radiation in the outer detector.
FC events are further divided into sub--GeV events ($E_{vis}\le
1.33$~GeV) and multi--GeV events ($E_{vis}\ge 1.33$~GeV).

In the Super--Ka\-mio\-kan\-de experiment 
for the double ratio $R$, from FC events (736 days) and PC events
(685 days), the following values were found: 
\bea
R&=& 0.67\pm 0.02\pm 0.05\qquad{\mathrm (sub-GeV)}
\label{R1}\\
R&=& 0.66\pm 0.04\pm 0.08\qquad{\mathrm (multi-GeV)}.
\label{R2}
\eea

Analogous results were obtained in the previous water
Cherenkov Ka\-mio\-kan\-de\cite{Kamioka} and IMB\cite{IMB} experiments
and in the iron calorimeter Soudan2\cite{Soudan} experiment:
\bea
R &=& 0.65\pm 0.05\pm 0.08 \qquad{\mathrm (Kamiokande)}
\nonumber\\
R &=& 0.54\pm 0.05\pm 0.11 \qquad{\mathrm (IMB)}
\label{oldratio}\\
R &=& 0.64\pm 0.11\pm 0.06 \qquad{\mathrm (Soudan2)}
\nonumber
\eea

The fact that $R$ is significantly less than one could imply
disappearance of atmospheric $\nu_{\mu}$ or appearence of
$\nu_e$ (or both). In the Super--Ka\-mio\-kan\-de experiment it was found
compelling evidence in favour of the disappearance of $\nu_{\mu}$ due
to neutrino oscillations:  in this experiment  a 
significant up--down asymmetry of the multi--GeV muon events was
discovered.

For atmospheric neutrinos the distances between production points
and detector can differ from $L\simeq 10$~Km (down--going neutrinos, 
$\theta=0$, $\theta$ being the zenith angle) to $L\simeq 10^4$~Km 
(up--going neutrinos, $\theta=\pi$). At high energies the effect 
of the Earth magnetic field is small and the expected number of 
neutrino events does not depend on the zenith angle $\theta$.
However the Super--Ka\-mio\-kan\-de collaboration found a significant 
$\theta$--dependence of the multi--GeV muon--neutrino events. For 
the integral up--down asymmetry
\beq
{\cal A}=\frac{ {\cal U}-{\cal D} }{ {\cal U} + {\cal D} }
\label{asym}
\eeq
it was obtained the value
\beq
{\cal A}_{\mu} = -0.311\pm 0.043\pm 0.010\, .
\label{asymmuon}
\eeq
Here ${\cal U}$ is the number of up--going events 
($-1\le\cos\theta\le -0.2$) and ${\cal D}$ is the number of 
down--going events ($0.2\le\cos\theta\le 1$). No significant 
asymmetry of the electron--neutrino events was found:
\beq
{\cal A}_e = -0.036\pm 0.067\pm 0.02\, .
\label{asymelec}
\eeq

The Super--Kamio\-kan\-de data can be described if we assume that 
$\nu_{\mu}\to\nu_{\tau}$ or $\nu_{\mu}\to\nu_s$ oscillations take place. 
In the $\nu_{\mu}\to\nu_{\tau}$ case the following best--fit values for
the oscillation parameters were found\cite{sam6}:
\beq
\sin^2{2\theta}=1,\qquad \Delta m^2=3.5\times 10^{-3}{\mathrm eV}^2
\label{bestfit1}
\eeq
(with $\chi^2_{\mathrm min}=6.21$ for 6.7 d.o.f.). In the case of 
$\nu_{\mu}\to\nu_s$  oscillations the best--fit values of the 
parameters are
\beq
\sin^2{2\theta}=1,\qquad \Delta m^2=4.5\times 10^{-3}{\mathrm eV}^2
\label{bestfit2}
\eeq
(with $\chi^2_{\mathrm min}=64.3$ for 67 d.o.f.).

\subsection{Solar neutrinos}

The energy of the sun is produced in the reactions of thermonuclear 
$pp$ and CNO cycles. From the thermodynamical point of view the 
energy of the sun is produced in the transition
\beq
2 e^- + 4p\longrightarrow ^4{\mathrm He} +2\nu_e\, .
\label{solarnu}
\eeq
Thus the production of energy in the sun is accompanied by the 
emission of electron neutrinos.

The main sources of solar neutrinos are the reactions of the $pp$ 
cycle listed in Table I.
As it is seen from the table, solar neutrinos are mainly low energy
$pp$ neutrinos and intermediate energy, monochromatic $^7$Be neutrinos,
while the high energy part of the solar neutrino spectrum is due to
the $^8$B decay.

\begin{table}[t]
\begin{center}
\begin{tabular}{ccc}
Reaction
&
Neutrino
&
Expected\\
& energy
& flux (cm$^{-2}$ s$^{-1}$)\\
& & BP 98\cite{BP98}\\
\hline
\hline
$p p \to d e^+ \nu_e$
&
$ \le 0.42$~MeV 
&
$6\cdot 10^{10}$\\
$e^{-} {^7{\mathrm Be}} \to \nu_e ^7{\mathrm Li}$
& ~0.86~MeV & $4.8\cdot 10^9$\\
$^8$B $\to ^8$Be$^* e^+\nu_e$
&$\le ~15$~MeV & $5\cdot 10^6$\\
\hline
\hline
\end{tabular}
\end{center}
%\refstepcounter{tables}
%\label{Table I}
\begin{center}
\footnotesize
Table I. Main sources of solar neutrinos.
\end{center}
\end{table}

The total flux of solar neutrinos is connected with the luminosity
of the sun, $L_{\odot}$ by the relation
\beq
Q\sum_{i=pp,\dots}\left(1-2\frac{{\bar E}_i}{Q}\right)\Phi_i=
\frac{L_{\odot}}{2\pi R^2}\, ,
\label{flux}
\eeq
where $Q=4m_p+2m_e-m_{^4{\mathrm He}}\simeq 26.7$~MeV is the energy
release in the transition (\ref{solarnu}), $\Phi_i$ is the total flux
of neutrinos from the source $i$ ($i=pp$, $^7$Be, $^8$B,$\dots$), $R$ is the
Sun--Earth distance and ${\bar E}_i$ is the average energy of neutrinos
from the source $i$. The above relation was derived under the assumption
that $P(\nu_e\to\nu_e)=1$. Notice that for $pp$ and $^7$Be neutrinos the
term ${\bar E}_i/Q$ can be neglected, while $\Phi_{^8{\mathrm B}}$ gives
a very small contribution to the left--hand side of (\ref{flux}).

\begin{table}[t]
\begin{center}
\begin{tabular}{cccc}
Experiment
&
{\begin{tabular}{c}
Observed\\
rate
\end{tabular}}
&
{\begin{tabular}{c}
Predicted\\
rate BP 98
\end{tabular}}
& 
$\frac{\displaystyle\mathrm data}{\displaystyle\mathrm prediction}$
\\
\hline
\hline
\\
Homestake\cite{Homesta}
& $2.56\pm 0.16\pm 0.11$ SNU
& $7.7^{+1.2}_{-1.0}$ SNU
& $0.33\pm 0.06$
\\
GALLEX\cite{Gallex}
& $77.5\pm 6.2^{+4.3}_{-4.7}$ SNU
& $129^{+8}_{-6}$ SNU
& $0.60\pm 0.07$
\\
SAGE\cite{Sage}
& $66.6^{+6.8+3.8}_{-7.1-4.0}$ SNU
& $129^{+8}_{-6}$ SNU
& $0.52\pm 0.07$
\\
Kamiokande\cite{Kamioka}
& 
{\begin{tabular}{c}
$(2.80\pm 0.19\pm 0.33)\cdot 10^6$
\\
(cm$^{-2}$s$^{-1}$)
\end{tabular}}
& 
{\begin{tabular}{c}
$(5.15^{+1.0}_{-0.7})\cdot 10^6$
\\
(cm$^{-2}$s$^{-1}$)
\end{tabular}}
& $0.54\pm 0.07$
\\
Super--Kamiokande\cite{Skamioka}
& 
{\begin{tabular}{c}
$(2.44\pm 0.05^{+0.09}_{-0.07})\cdot 10^6$
\\
(cm$^{-2}$s$^{-1}$)
\end{tabular}}
& 
{\begin{tabular}{c}
$(5.15^{+1.0}_{-0.7})\cdot 10^6$
\\
(cm$^{-2}$s$^{-1}$)
\end{tabular}}
& $0.47^{+0.07}_{-0.09}$
\\
\\
\hline
\hline
\\
\end{tabular}
\end{center}
%\refstepcounter{tables}
%\label{Table I}
\begin{center}
\footnotesize
Table II. Results of solar neutrino experiments 
(1~SNU=$10^{-36}$ events/(atom~sec).
\end{center}
\end{table}

The results of five solar neutrino experiments are available at present:
they are reported in Table II.

In the radiochemical Homestake experiment\cite{Homesta} solar $\nu_e$'s
are detected by observing the Pontecorvo--Davis reaction
\beq
\nu_e+^{37}{\mathrm Cl}\longrightarrow e^- + ^{37}{\mathrm Ar}.
\label{PonDavis}
\eeq
The threshold of this process is $E_{th}=0.81$~MeV. Thus mainly $^8$B
and $^7$Be are detected in this experiment (according to the Standard
Solar Model (SSM) the
contributions of $^8$B and $^7$Be neutrinos to the total rate 
are $77\%$ and $14\%$, respectively).

In the radiochemical GALLEX\cite{Gallex} and SAGE\cite{Sage} experiments
solar $\nu_e$'s are detected through the observation of the reaction
\beq
\nu_e + ^{71}{\mathrm Ga}\longrightarrow e^- + ^{71}{\mathrm Ge}\, ,
\label{Gallium}
\eeq
whose threshold is $E_{th}=0.23$~MeV. Hence neutrinos from all sources
can be detected by these experiments. According to the SSM, the
contributions of $pp$, $^7$Be and $^8$B neutrinos to the total rate
are, respectively, $54\%$, $27\%$, $10\%$.

Finally in the direct counting Kamio\-kan\-de\cite{Kamioka2} and
Super--Kamio\-kan\-de\cite{Skamioka} experiments, 
solar neutrinos are detected by
the observation of recoil electrons from the process
\beq
\nu_e + e\longrightarrow \nu_e + e\, .
\label{NCnue}
\eeq
Due to the high energy threshold ($E_{th}=7$~MeV in the Kamio\-kan\-de
experiment; in the Super--Ka\-mio\-kan\-de one $E_{th}=6.5$~MeV and in 
the most recent runs $E_{th}=5.5$~MeV)
 only $^8$B neutrinos are detected in these experiments.

As it is seen from Table II, the observed rates in all solar neutrino
experiments are significantly smaller than the predicted rates.
The existing data cannot be described if we assume that
$P(\nu_e\to\nu_e)=1$, even when the total fluxes $\Phi_i$ are left as
free fitting parameters (there are no acceptable fits at $99.99\%$ 
CL\cite{sam9} ). 

Instead, the data presented in Table II can be accounted for by
assuming that there is a two--neutrino mixing (if the SSM
values for the neutrino fuxes $\Phi_i$ are used). In the case of
$\nu_e\to\nu_\mu$ (or $\nu_\tau$) transitions the following best--fit
values of the oscillation parameters were obtained\cite{sam6}:
\beq
 \sin^2 2\theta=5\cdot 10^{-3},\quad
\Delta m^2=7.1\cdot 10^{-6}~{\mathrm eV}^2;\,\, CL=1.6\%
\label{SMA}
\eeq
\begin{center}
(small mixing angle MSW solution)
\end{center}
\beq
 \sin^2 2\theta=0.7,\quad
\Delta m^2=2.8\cdot 10^{-5}~{\mathrm eV}^2;\,\, CL=1.2\%
\label{LMA}
\eeq
\begin{center}
(large mixing angle MSW solution) 
\end{center}
\beq
 \sin^2 2\theta=0.89,\quad
\Delta m^2=4.3\cdot 10^{-10}~{\mathrm eV}^2;\,\, CL=9.9\%
\label{VO}
\eeq
\begin{center}
(vacuum oscillation solution)
\end{center}

These solutions were obtained by fitting the data, presented in
Table II, together with the Super--Ka\-mio\-kan\-de data on the
measurement of the spectrum of recoil electrons in $\nu_e\to\nu_e$
scattering.

In the nearest future two new solar neutrino experiments, SNO\cite{SNO}
and BOREXINO\cite{Borexino}, will be started. In the SNO experiment (heavy
water Cherenkov detector, 1~Kton of $D_2O$) solar neutrinos will be
detected by measuring the CC process
\beq
\nu_e\, d \longrightarrow e^-\, p p
\label{CCSNO}
\eeq
as well as the NC one:
\beq
\nu\, d \longrightarrow  \nu\,n\, p
\label{NCSNO1}
\eeq
and the process
\beq
 \nu\, e\longrightarrow \nu_e\, e
\label{NCSNO2}
\eeq
Due to the high energy threshold ($E_{th}=5$~MeV for the processes
(\ref{CCSNO}) and (\ref{NCSNO2}) and $E_{th}=2.2$~MeV for the NC
process (\ref{NCSNO1}) ), mainly $^8$B neutrinos will be detected in this
experiment.

By measuring the electron energy in the CC process (\ref{CCSNO})
it will be possible to determine the spectrum of the solar $\nu_e$'s
on the Earth. Moreover the detection of solar neutrinos through the
observation of the NC process (\ref{NCSNO1}) (the neutron will be 
detected) will allow to obtain information on the flux of all active
neutrinos, $\nu_e, \nu_{\mu}, \nu_{\tau}$ on the Earth. From the
comparison of these data a model independent conclusion on the
transitions of solar $\nu_e$'s into other states can be drawn.

In the BOREXINO experiment\cite{Borexino} (300~tons of liquid scintillator) 
the observation of the proces $\nu e\to \nu e$ will allow to detect 
monochromatic $^8$B neutrinos, with energy $E=0.86$~MeV. The threshold 
for the detection of the recoil electrons is $E_{th}=250$~keV. 
The SSM predicts for this experiment 
$\simeq 50$ events/day. In the case of vacuum oscillations, a
significant seasonal variation of the number of events would be
observed.

\subsection{LSND experiment}

Only in one accelerator neutrino experiment (LSND\cite{LSND})
 indications in favour of neutrino ascillations were found. This experiment 
was done at the Los Alamos  linear accelerator (the proton energy being
 800~MeV). Neutrinos were produced in the decays of $\pi^+$ and 
$\mu^+$ at rest:
\beq
\begin{array}{cc}
\pi^+ &\longrightarrow \mu^+ \nu_\mu\\
\mu^+ &\longrightarrow e^+\nu_e{\bar\nu}_\mu
\end{array}
\label{pion}
\eeq
The Large Scintillator Neutrino Detector (LSND) was located
at a distance of about 30~m from the neutrino source. The LSND
collaboration was searching for ${\bar\nu}_e$ through the
observation of the process
\beq
{\bar\nu}_e + p\longrightarrow e^+ + n.
\label{epiu}
\eeq
Both $e^+$ and delayed $\gamma$'s from the capture $np\to d\gamma$ 
were detected.

In the LSND experiment $33.9\pm 8.0$ events were observed in the interval 
of $e^+$ energies $30\le E\le 60$~MeV. If these events are due to 
${\bar\nu}_\mu\to{\bar\nu}_e$ oscillations, for the transition 
probability it was found:
\beq
P({\bar\nu}_\mu\to{\bar\nu}_e) = (0.31\pm 0.09\pm 0.06)\cdot 10^{-2}\, .
\label{pmue}
\eeq
It corresponds to the following allowed ranges of the oscillation 
parameters:
\beq
\begin{array}{ccc}
0.3 &< \Delta m^2 &\le 1 {\mathrm eV}^2\\
2\cdot 10^{-3} &\le \sin^2 {2\theta} &\le 4\cdot 10^{-2}\, .
\end{array}
\label{param}
\eeq
These values do not contradict the data of other neutrino oscillation
experiments, including those of the KARMEN collaboration\cite{Karmen},
searching for ${\bar\nu}_\mu\to{\bar\nu}_e$ transition at the spallation 
neutron facility of the Rutherford Laboratory. The LSND region of the 
oscillation parameters will be thoroughly investigated by the future 
BOONE\cite{Boone} experiment at Fermilab.

\section{Neutrino masses and mixing from oscillation data}

From all existing neutrino oscillation data, it follows that there are
three different scales of $\Delta m^2$: hence in order 
to describe these data we must
assume the existence of at least four massive neutrinos. If the data
of the LSND experiment should not be confirmed by future experiments, then
it would be enough to assume the existence of three massive neutrino only.
We will consider both these possibilities.

\subsection{Mixing of three massive neutrinos}

For the mixing of three massive neutrinos it is natural to assume the 
following mass hierarchy:
\[m_1 \ll m_2 \ll m_3 \]
with $\Delta m_{21}^2$ and $\Delta m_{31}^2$  
relevant  for the oscillations of solar and atmospheric
neutrinos, respectively.

It is easy to see that for atmospheric and LBL neutrino oscillation
experiments the inequality
\beq
\frac{\Delta m_{21}^2 L}{2p}\ll 1
\label{inequa}
\eeq
holds. Hence, from (\ref{prob1}), the $\nu_\alpha\to\nu_\beta$ transition 
probability becomes\cite{sam11}:
\beq
P\left(\nu_\alpha\to\nu_\beta\right) =\frac{1}{2} A_{\alpha;\beta}
\left(1-\cos\frac{\Delta m_{31}^2 L}{2p}\right), 
\qquad(\alpha\ne\beta);
\label{Pab}
\eeq
and the  $\nu_{\alpha}$ survival probability  is given by:
\bea
P\left(\nu_\alpha\to\nu_\alpha\right)&=&1-\sum_{\beta\ne\alpha}
P\left(\nu_\alpha\to\nu_\beta\right)
\nonumber\\
&=& 1-\frac{1}{2}B_{\alpha;\alpha}
\left(1-\cos\frac{\Delta m_{31}^2 L}{2p}\right),
\label{Paa}
\eea
 the oscillation amplitudes being 
\beq
\begin{array}{cl}
A_{\alpha;\beta}&= 4|U_{\alpha 3}|^2\, |U_{\beta 3}|^2\\
B_{\alpha;\alpha} &= 4|U_{\alpha 3}|^2(1- |U_{\alpha 3}|^2).
\end{array}
\label{trpro}
\eeq
The expressions (\ref{Pab}), (\ref{Paa}) have the same form as the ones
for two--neutrino transition probabilities, eqs. (\ref{trans1}),
(\ref{trans2}).
They describe, however, all possible transitions between three flavour 
neutrinos.

Notice that, due to the unitarity of the mixing matrix, 
$\sum_{\alpha=e,\mu,\tau}|U_{\alpha 3}|^2=1$. Thus, in the framework of 
three--neutrino mixing, the transition probabilities for the atmospheric
and LBL neutrino experiments are described by three parameters: 
$|U_{e 3}|^2$, $|U_{\mu 3}|^2$ and $\Delta m_{31}^2$.

Information on the parameter $|U_{e3}|^2$ can be obtained from the LBL
experiment CHOOZ\cite{CHOOZ}. In this first reactor LBL experiment (the
distance between the reactor and the detector being about 1~Km) there
were found no indications in favour of neutrino oscillations. The ratio
$R$ between the number of observed events and the number of predicted
events (under the assumption that there are no oscillations) was found
to be:
\beq
R=0.98\pm 0.04\pm 0.04 .
\label{ratiochooz}
\eeq
From the results of the experiment the CHOOZ collaboration extracted the
exclusion plot in the plane of parameters $(\sin^2{2\theta}, \Delta m^2)$.
It was shown that the region
\beq
\begin{array}{ccc}
&&0.2\le \sin^2{2\theta}_{\mathrm CHOOZ} \le 1\\
 &&\qquad\Delta m_{31}^2 \ge 2\cdot 10^{-3}{\mathrm eV}^2
\end{array}
\label{exclchooz}
\eeq
is excluded.

From eq.(\ref{trpro}) it is easy to see that the parameter
$|U_{e3}|^2$ is connected with the amplitude $B_{e;e}$ by the
relation
\beq
|U_{e3}|^2=\frac{1}{2}\left(1\pm \sqrt{1-B_{e;e}}\right)\, .
\label{Ue3}
\eeq
From the exclusion curve of the CHOOZ experiment in the region
$\Delta m_{31}^2\ge 2\cdot 10^{-3}$~eV$^2$ the following upper bound
can be obtained
\beq
B_{e;e}\le B^0_{e;e},
\label{B0ee}
\eeq
the values $B^0_{e;e}\equiv(\sin^2{2\theta})_{\mathrm CHOOZ}$ depending on
$\Delta m_{31}^2$. From (\ref{Ue3}) and (\ref{B0ee}) the parameter
$|U_{e3}|^2$ turns out to be subject to the conditions\cite{sam12}
\beq
|U_{e3}|^2\le a_0\qquad{\mathrm or}\qquad
|U_{e3}|^2\ge 1-a_e^0,
\label{Ue3bis}
\eeq
where
\beq
a_e^0=\frac{1}{2}\left(1-\sqrt{1-B_{e;e}^0}\right)\, .
\eeq
At $\Delta m_{31}^2\ge 2\cdot 10^{-3}$~eV$^2$ the CHOOZ exclusion curve
entails $B^0_{e;e}<0.18$ and hence
\beq
|U_{e3}|^2\le 5\cdot 10^{-2}\qquad{\mathrm or}
\qquad |U_{e3}|^2\ge 0.95
\eeq

Large values of $|U_{e3}|^2$ are excluded by the results of solar
neutrino experiments. Indeed for the probability of solar neutrinos
to survive we have\cite{sam13}
\beq
P^{\mathrm solar}(\nu_e\to\nu_e)=\left(1-|U_{e3}|^2\right)^2
P^{(1,2)}(\nu_e\to\nu_e) +|U_{e3}|^4
\label{Psolar}
\eeq
where $P^{(1,2})(\nu_e\to\nu_e)$ is the transition probability due
to the coupling of $\nu_e$ to $\nu_1,\nu_2$. If $|U_{e3}|^2\ge 0.95$
then $P^{\mathrm solar}(\nu_e\to\nu_e)\ge 0.90$, a result which is not
compatible with the outcome of solar neutrino experiments. Hence we
come to the conclusion that
\beq
|U_{e3}|^2\le 5\cdot 10^{-2}.
\label{noname}
\eeq

From the data of atmospheric neutrino experiments it also follows that
the element $|U_{e3}|^2$ is small in the region $10^{-3}\le \Delta m_{31}^2
\le 8 \cdot 10^{-3}$~eV$^2$: in these experiments there is no indication 
in favour of $\nu_{\mu}\to\nu_e$ oscillations; moreover the amplitude 
of $\nu_{\mu}\to\nu_{\tau}$ 
oscillations is close to the maximum allowed value, which means that
$|U_{\mu 3}|^2\simeq|U_{\tau 3}|^2\simeq 1/2$ and $|U_{e3}|^2\simeq 0$.

The unitary $3\times 3$ matrix $U$ is characterized by three mixing angles,
$\theta_{12},\theta_{13},\theta_{23}$, and one phase, $\phi$. The condition
$U_{e3}\simeq 0$ is equivalent to $\theta_{13}\simeq 0$; in this case the 
phase is irrelevant and the remaining angles $\theta_{12}$ and $\theta_{23}$
can be determined from solar and atmospheric neutrino data, respectively
(the oscillations of solar and atmospheric neutrinos are 
decoupled\cite{sam12}).

\subsection{Mixing of four massive neutrinos}

Two types of mass spectra are possible in the presence of three scales
for $\Delta m^2$ ($\Delta m_{\mathrm atm}^2\simeq 10^{-3}$~eV$^2$,
$\Delta m_{\mathrm solar}^2\simeq 10^{-5}$~eV$^2$ ($10^{-10}$~eV$^2$),
$\Delta m_{\mathrm LSND}^2\simeq 1$~eV$^2$). In the spectra of the first 
type a group of three close masses is separated from the fourth one by the
``LSND gap'' of about 1~eV. In the spectra of the second type two pairs 
of close masses are separated by $\simeq 1$~eV gap.

Neutrino mass spectra of the first type are not compatible with 
the data of neutrino oscillation experiments\cite{sam16},\cite{sam17}.
In fact let us consider the case of a mass hierarchy of four neutrinos.
In SBL experiments $\Delta m_{21}^2 L/2p\ll 1$ and 
$\Delta m_{31}^2 L/2p\ll 1$; in this situation, from equation 
(\ref{prob1}) for the $\nu_{\alpha}\to\nu_{\beta}$ transition probability, 
we can obtain expressions similar to (\ref{Pab})--(\ref{trpro}), providing the 
following replacements are performed: $U_{\alpha 3}\to U_{\alpha 4}$, 
$\Delta m_{31}^2\to \Delta m_{41}^2$. Furthermore, taking into account 
solar and atmospheric neutrino data, we have, in analogy with (\ref{noname}),
\beq
|U_{\alpha 4}|^2\le a^0_{\alpha}\qquad (\alpha=e,\mu).
\label{ineq}
\eeq
Here 
\beq
a^0_{\alpha}=\frac{1}{2}\left(1-\sqrt{1-B^0_{\alpha;\alpha}}\right)
\eeq
and $B^0_{\alpha;\alpha}$ can be found from the exclusion curves of SBL 
reactor and accelerator disappearence experiments. We get 
\beq
\begin{array}{cccc}
a^0_e &\le 4\cdot 10^{-2}\qquad &{\mathrm for}\qquad 
&\Delta m_{41}^2\ge 0.1{\mathrm eV}^2\\
a^0_{\mu} &\le 2\cdot 10^{-1}\qquad &{\mathrm for}\qquad 
&\Delta m_{41}^2\ge 0.3{\mathrm eV}^2
\end{array}
\eeq
For the amplitude of SBL $\nu_{\mu}\to\nu_e$ transitions we have the
following upper bound:
\beq
A_{\mu;e}\equiv 4|U_{e4}|^2 |U_{\mu 4}|^2\le 4a^0_e a^0_{\mu}.
\label{upperb}
\eeq
However it can be shown\cite{sam16},\cite{sam18} that the quantity 
$4a^0_e a^0_{\mu}$ is too small to be 
compatible with the allowed region in the plot obtained by the LSND 
collaboration. The same considerations apply to all other spectra of the
first type.

%%%%%%%%%%%%%%%%%%%%%%%  Fig.1   %%%%%%%%%%%%%%%%%%%%%%%%%%%%%%%%%%%
\begin{eqnarray}
\mbox{(A)} &&
\quad
\underbrace{
\overbrace{m_1 < m_2}^{\mathrm{atm}}
\ll
\overbrace{m_3 < m_4}^{\mathrm{solar}}
}_{\mathrm{LSND}} \;, \nonumber \\
%\qquad \mbox{and} \qquad
\mbox{(B)} &&
\quad
\underbrace{
\overbrace{m_1 < m_2}^{\mathrm{solar}}
\ll
\overbrace{m_3 < m_4}^{\mathrm{atm}}
}_{\mathrm{LSND}}
\nonumber
\end{eqnarray}

\begin{center}
Fig.~1 
{\small The schemes A and B for the  neutrino mass spectrum discussed
in the text.}
\end{center}

%%%%%%%%%%%%%%%%%%%%%%%%%%%%%%%%%%%%%%%%%%%%%%%%%%%%%%%%%%%%%%%%%%%%

Only two schemes of mixing of four massive neutrinos (A and B), with the 
mass spectra shown in Fig.~1, can describe all existing 
neutrino oscillation data. In these schemes, in place of the inequality 
(\ref{ineq}), we have:
\beq
\begin{array}{ccccc}
A. &\sum_{i=1,2}|U_{ei}|^2 &\le a_e^0\qquad
B. &\sum_{k=3,4}|U_{ek}|^2 &\le a_e^0\\
 &\sum_{k=3,4}|U_{\mu k}|^2 &\le a_{\mu}^0\qquad
 &\sum_{i=1,2}|U_{\mu i}|^2 &\le a_{\mu}^0
 \end{array}
\label{masscheme}
\eeq
In both schemes, for the amplitude of SBL $\nu_\mu\to\nu_e$ transition we have
the following upper bound
\beq
A_{\mu;e}\le 4\sum_{i=1,2}|U_{ei}|^2\cdot\sum_{i=1,2}|U_{\mu i}|^2
\le 4{\mathrm min}\left(a^0_e, a^0_{\mu}\right)\, .
\label{ampli}
\eeq
This bound is linear in the (small) quantities $a^0_e$ and $a^0_\mu$
and is compatible with the LSND results.

The above schemes of mixing of four neutrinos suggests the existence
of a sterile neutrino. Taking into account the Big Bang Nucleosynthesis
constraint on the effective number of neutrinos, it can be
shown\cite{sam19},\cite{sam17}
that in both schemes A and B the dominant transition of solar neutrinos is
the $\nu_e\to\nu_s$ one and the dominant transition of atmospheric
neutrinos is $\nu_\mu\to\nu_\tau$. These predictions will be tested
by future solar, atmospheric and LBL experiments.

\section{Conclusions}

The latest discoveries opened a new era in neutrino physics: massive
and mixed neutrinos became real physical objects. Many new experiments
must be implemented to investigate further neutrino properties and to
reveal the physics which governs them. Undoubtedly the investigation of
neutrino properties is one of the most important directions in the
search for new physics.

Neutrino physics could also greatly help in solving other problems.
One example is the so--called problem of the spin of the nucleon, which is
connected with the strange content  of the nucleon. The detailed
investigation of NC neutrino induced processes and, specifically,
the elastic neutrino (antineutrino)--proton scattering could allow
to obtain model independent information on the strange form factors
of the nucleon\cite{sam20}. Here the neutrino plays the role of a 
probe to test the hadronic structure at a high level of precision: 
for this a deep knowledge of the structure of the neutrino 
currents is required.
\vskip 1cm

{\bf Acknowledgements}: 
We would like to warmly thank C. Giunti and W. Grimus for numerous fruitful
discussions on the problem of neutrinos and for useful advices in the 
preparation of the manuscript. S.M.B. would like to thank Prof. H. 
Pietschmann, 
Director of the Institute for Theoretical Physics of the University of
Vienna, for the hospitality and support.


\small

\begin{thebibliography}{99}

\bibitem{SKamio98} 
T.~Kajita, Super-Kamiokande Coll., Talk presented at \textit{Neutrino '98},
Takayama, Japan, June 1998;\\
Y.~Fukuda \textit{et al.}, Super-Kamiokande Coll., 
Phys. Rev. Lett. \textbf{81}, 1562 (1998).

\bibitem{IMB} 
R.~Becker-Szendy \textit{et al.}, IMB Coll., 
Nucl. Phys. B (Proc. Suppl.) \textbf{38}, 331 (1995).

\bibitem{Soudan} 
W.W.M. Allison \textit{et al.}, Soudan-2 Coll., 
Phys. Lett. B \textbf{391}, 491 (1997).

\bibitem{Kamioka} 
Y.~Fukuda \textit{et al.},
Kamiokande Coll., Phys. Lett. B \textbf{335}, 237 (1994).

\bibitem{Homesta} 
R.~Davis, Harmer and K.C. Hoffman,
Phys. Rev. Lett. \textbf{21}, 1205 (1968);\\
R.~Davis, Prog. Part. Nucl. Phys. \textbf{32}, 13 (1994);\\
B.T. Cleveland \textit{et al.},
Astrophys. J. \textbf{496}, 505 (1998).

\bibitem{Gallex} 
P.~Anselmann \textit{et al.}, GALLEX Coll., 
Phys. Lett. B \textbf{285}, 376 (1992);\\
W.~Hampel \textit{et al.}, GALLEX Coll., 
Phys. Lett. B \textbf{388}, 384 (1996).

\bibitem{Sage} 
A.I. Abazov \textit{et al.}, SAGE Coll., 
Phys. Rev. Lett. \textbf{67}, 3332 (1991);\\
D.N. Abdurashitov \textit{et al.}, SAGE Coll., 
Phys. Rev. Lett. \textbf{77}, 4708 (1996).

\bibitem{Kamioka2} 
K.S. Hirata \textit{et al.}, Kamiokande Coll., 
Phys. Rev. D \textbf{38}, 448 (1988);\\
K.S. Hirata \textit{et al.}, Kamiokande Coll., 
Phys. Rev. Lett. \textbf{77}, 1683 (1996).

\bibitem{Skamioka} 
K.~Inoue, Super-Kamiokande Coll., 
Talk presented at the $5^{\mathrm{th}}$ International Workshop on 
\textit{Topics in Astroparticle and Underground Physics}, Gran Sasso, 
Italy, September, 1997;\\
Y.~Fukuda \textit{et al.}, Super-Kamiokande Coll., 
Phys. Rev. Lett. \textbf{81}, 1158 (1998);\\
Y.~Suzuki, Super-Kamiokande Coll., 
Talk presented at \textit{Neutrino '98}, 1998.

\bibitem{MACRO} 
M.~Ambrosio \textit{et al.}, MACRO Coll., 
Phys. Lett. B \textbf{434}, 451 (1998).

\bibitem{LSND} 
C.~Athanassopoulos \textit{et al.}, LSND Coll., 
Phys. Rev. Lett. \textbf{77}, 3082 (1996);\\
C.~Athanassopoulos \textit{et al.}, LSND Coll., 
Phys. Rev. Lett. \textbf{81}, 1774 (1998);\\
C.~Athanassopoulos \textit{et al.}, LSND Coll.,
Phys. Rev. Lett. \textbf{75}, 2650 (1995);\\
C.~Athanassopoulos \textit{et al.}, LSND Coll.,
Phys. Rev. C \textbf{54}, 2685 (1996).

\bibitem{Minos}
D.~Ayres \textit{et al.},
MINOS Coll., report NUMI-L-63 (1995).

\bibitem{Icarus}
P.~Cennini \textit{et al.}, ICARUS Coll.,
report LNGS-94/99-I (1994).

\bibitem{Opera}
H.~Shibuya \textit{et al.}, OPERA Coll.,
report CERN-SPSC-97-24 (1997);\\
P.~Picchi and F.~Pietropaolo,
Talk presented at \textit{Neutrino '98}, preprint hep-ph/9812222 (1998).
                                        
\bibitem{SNO}
A.~McDonald, SNO Coll.,
Talk presented at \textit{Neutrino '98}, 1998;\\
SNO WWW page: http://\-www.\-sno.\-queensu.\-ca.

\bibitem{Borexino}
L.~Oberauer, Borexino Coll.,
Talk presented at \textit{Neutrino '98}, 1998;\\
Borexino WWW pages: http://\-almime.\-mi.\-infn.\-it/;
  http://\-pupgg.\-princeton.\-edu/\~{}borexino/\-welcome.\-html.

\bibitem{Kamland}
F.~Suekane,
preprint TOHOKU-HEP-97-02 (1997);\\
A.~Suzuki,
Talk presented at NOW'98, Amsterdam, September 1998.

\bibitem{Boone}
Booster Neutrino Experiment (BooNE), http://\-nu1.lampf.lanl.gov/\-BooNE.

\bibitem{sam1}
F.~Wilczek,
Talk presented at \textit{Neutrino '98}, 1998
(hep-ph/9809509).

\bibitem{LEP} 
C.~Caso \textit{et al.}, Particle Data Group, 
Eur. Phys. J. C \textbf{3}, 1 (1998).

\bibitem{CSW} S.M. Bilenky, C. Giunti and W. Grimus, 
hep-ph/9812360, March 1999, to be published in Progress in Particle 
and Nuclear Physics.

\bibitem{sam3} 
M.~Gell-Mann, P.~Ramond and R.~Slansky,
in \textit{Supergravity}, p.~315, edited by F. van Nieuwenhuizen and D.
Freedman, North Holland, Amsterdam, 1979;\\
T.~Yanagida, 
Proc. of the \textit{Workshop on Unified Theory and the Baryon Number 
of the Universe}, KEK, Japan, 1979;\\
R.N. Mohapatra and G.~Senjanovi{\'c},
Phys. Rev. Lett. \textbf{44}, 912 (1980).

\bibitem{sam4} 
B.~Pontecorvo,
J. Exptl. Theoret. Phys. \textbf{34}, 247 (1958)
[Sov. Phys. JETP \textbf{7}, 172 (1958)];\\
B.~Pontecorvo,
Zh. Eksp. Teor. Fiz. \textbf{53}, 1717 (1967)
[Sov. Phys. JETP \textbf{26}, 984 (1968)]; \\
V.~Gribov and B.~Pontecorvo,
Phys. Lett. B \textbf{28}, 493 (1969);\\
S.M. Bilenky and B.~Pontecorvo,
Phys. Rep. \textbf{41}, 225 (1978).

\bibitem{sam5}
W.~Grimus, P.~Stockinger and S.~Mohanty,
Phys. Rev. D \textbf{59}, 013011 (1998).

\bibitem{sam6}
Y. Suzuki, 
in Proc. of {\it WIN 99}, Capetown, January 24-30, 1999.

\bibitem{sam9}
J.N. Bahcall, P.I. Krastev and A.Yu. Smirnov,
Phys. Rev. D \textbf{58}, 096016 (1998).

\bibitem{BP98}
J.N. Bahcall, S.~Basu and M.H. Pinsonneault,
Phys. Lett. B \textbf{433}, 1 (1998).

\bibitem{Karmen}
B.~Zeitnitz, KARMEN Coll.,
Talk presented at \textit{Neutrino '98}, 1998.

\bibitem{sam11}
A.~De Rujula \textit{et al.},
Nucl. Phys. B \textbf{168}, 54 (1980);\\
V.~Barger and K.~Whisnant,
Phys. Lett. B \textbf{209}, 365 (1988);\\
S.M. Bilenky, M.~Fabbrichesi and S.T. Petcov,
Phys. Lett. B \textbf{276}, 223 (1992);\\
S.M. Bilenky, C.~Giunti and C.W. Kim,
Astrop. Phys. \textbf{4}, 241 (1996).

\bibitem{CHOOZ}
M. Apollonio \textit{et al.}, CHOOZ Coll., Phys. Lett. B \textbf{420}, 397;\\
C. Bemporad, Talk presented at the Ringberg Euroconference \textit{New
Trends in Neutrino Physics}, Ringberg Castle, Tegernsee, Germany, May, 1998.
         
\bibitem{sam12}
S.M. Bilenky and C.~Giunti,
Phys. Lett. B \textbf{444}, 379 (1998).
               
\bibitem{sam13}
X.~Shi and D.N. Schramm,
Phys. Lett. B \textbf{283}, 305 (1992).

\bibitem{sam16}
S.M. Bilenky, C.~Giunti and W.~Grimus,
Eur. Phys. J. C \textbf{1}, 247 (1998).

\bibitem{sam17}
N.~Okada and O.~Yasuda,
Int. J. Mod. Phys. A \textbf{12}, 3669 (1997).

\bibitem{sam18}
S.M. Bilenky, C. Giunti, W. Grimus, T. Schwetz,
hep-ph/9903454.

\bibitem{sam19}
S.M. Bilenky, C.~Giunti, W.~Grimus and T.~Schwetz,
preprint hep-ph/9804421 (1998)
[to be published in Astropart. Phys.].

\bibitem{sam20}
W.M. Alberico, S.M. Bilenky, C. Giunti and C. Maieron,
Z. f\"ur Physik \textbf{ C 70}, 463 (1996) and references therein.

\end{thebibliography}
\end{document}